\documentclass{emulateapj}

\usepackage{natbib}
\begin{document}

\title{Solar System Analogs Around \emph{IRAS}-Discovered Debris Disks}
\author{Christine H. Chen\altaffilmark{1}, Patrick Sheehan\altaffilmark{2},  
	      Dan M.  Watson\altaffilmark{2}, P. Manoj Puravankara\altaffilmark{2},
               Joan R.  Najita\altaffilmark{3}}
\altaffiltext{1}{Space Telescope Science Institute, 
	        3700 San Martin Dr.,
                 Baltimore, MD 21218; 
                 cchen@stsci.edu}
\altaffiltext{2}{Department of Physics and Astronomy, 
                 University of Rochester, 
                 Rochester, NY 14627}
\altaffiltext{3}{NOAO, 
	        950 North Cherry Avenue, 
                 Tucson, AZ 85726}

\begin{abstract}
We have rereduced \emph{Spitzer} IRS spectra and reanalyzed the SED's of three nearby debris disks: $\lambda$ Boo, HD 139664, and HR 8799. We find that that the thermal emission from these objects is well modeled using two single temperature black body components. For HR 8799 -- with no silicate emission features despite a relatively hot inner dust component ($T_{gr}$ = 150 K) -- we infer the presence of an asteroid belt interior to and a Kuiper Belt exterior to the recently discovered orbiting planets. For HD 139664, which has been imaged in scattered light, we infer the presence of strongly forward scattering grains, consistent with icy or porous grains, if the cold, outer disk component generates both the observed scattered light and thermal emission. Finally, careful analysis of the $\lambda$ Boo SED suggests that this system possesses a central clearing, indicating that selective accretion of solids onto the central star does not occur from a dusty disk.
\end{abstract}

\keywords{stars: circumstellar matter--- planetary systems: formation}

\section{Introduction}
Our solar system possesses four terrestrial planets, an asteroid belt, four Jovian planets, and a Kuiper belt. Whether this planetary architecture is common or rare has not yet been quantified. Dusty disks, believed to be generated by collisions between parent bodies, analogous to asteroid and Kuiper belt dust rings, have been detected around main sequence stars via thermal infrared emission. Space-based mid- and far-infared surveys indicate that $\sim$10-30\% of main sequence stars possess infrared excess (Lagrange et al. 2000) and that 95\% of these debris disks have color temperatures similar to that expected for cold dust in our Kuiper Belt (Meyer et al. 2007), suggesting that exo-Kuiper belts are common around main sequence stars. The remaining 5\% of debris disks possess massive, warm, dusty disks that may be the result of planet formation in self-stirred disks (Kenyon \& Bromley 2004) or massive, stochastic events such as the period of Late Heavy Bombardment in our solar system (Wyatt et al. 2007); however, whether lower mass, steady-state, asteroid belts are common or rare is not yet known. At the current time, only the exo-asteroid belt around $\zeta$ Lep has been spatially resolved (Moerchen et al. 2007).

Warm, asteroidal dust may be more prevalent than currently inferred from broad band photometric surveys. High resolution imaging suggests that some debris disks with cold spectral energy distribution (SED) color temperatures ($T_{gr}$ $<$ 100 K) may also possess warm components. Spatially resolved scattered light and thermal emission images of $\sim$15 resolved debris disks reveal two distinct dust architectures: (1) continuous disks with surface density distributions that are well approximated by a combination of a rising and a decaying power law (e.g. $\beta$ Pic, AU Mic) and (2) narrow belts with sharp inner edges (e.g. HR 4796A, Fomalhaut). Both geometries may possess dust at terrestrial temperatures. Ground-based mid-infrared imaging of $\beta$ Pic detected an unresolved component of infrared excess originating from distances $<$5 AU away from the central star (Telesco et al. 2005; Wahhaj et al. 2003) indicative of warm dust. Similar imaging of HR 4796A and Fomalhaut may also have detected unresolved warm dust components that are spatially distinct from the cold, narrow dust rings that are resolved in scattered light and thermal emission (Koerner et al. 1998; Stapelfeldt et al. 2004).

Unfortunately, the majority debris disks discovered thus far can not currently be spatially resolved because their surface brightnesses are too faint or their angular sizes are too small. Additional techniques may shed light on the presence of warm dust in cold debris systems: (1) \emph{Spitzer} 5.5- 40 $\mu$m SED observations may reveal the presence of an additional component of warm dust (this study) and (2) nulling interferometry observations may reveal an unresolved dust component close to the star (Liu et al. 2004). We have been conducting a \emph{Spitzer} IRS spectroscopic study of \emph{IRAS}-discovered debris disks (Chen et al. 2006). Rereduction and reanalysis of our previously-published observations finds four objects ($\lambda$ Boo, HR 5825 (HD 139664), HR 8799) with warm dust in addition to the cold dust that was originally discovered using \emph{IRAS}. The recent direct detection of three orbiting giant planets around HR 8799 (Marois et al. 2009) allows us to compare the planetary architecture of this system with that of our own. The synthesis of resolved scattered light images (Kalas et al. 2006) and the infrared SED allows us to place new constraints on the properties of dust grains around HD 139664.

\section{Observations}

The debris disks around the stars $\lambda$ Boo, HR 5825 (HD 139664), and HR 8799 were first identified based on the presence of 60 $\mu$m \emph{IRAS} excesses, 1 - 2 orders of magnitude brighter than expected from their stellar photospheres (Backman \& Paresce 1993). Recent \emph{Spitzer} MIPS 24 $\mu$m and 70 $\mu$m photometry has confirmed the presence of unresolved 24 $\mu$m and/or 70 $\mu$m excesses associated with $\lambda$ Boo (Su et al. 2006) and HD 139664 (Beichman et al. 2006). The 7$\arcsec$ and 21$\arcsec$ angular resolutions of \emph{Spitzer} at 24 and 70 $\mu$m, respectively, are a significant improvement from the 0.75$\arcmin$$\times$4.6$\arcmin$ and 1.5$\arcmin$$\times$4.7$\arcmin$ angular resolutions of \emph{IRAS} at 25 and 60 $\mu$m, respectively, and better exclude confusion from the background and/or other nearby sources. Chen et al. (2006) measured the shape of the hot dust continuum emitted from these objects from low resolution \emph{Spitzer} IRS spectra  and inferred the presence of warm dust with $T_{gr}$ = 100 K - 150 K, hotter than estimated from photometric observations alone. 

We present updated IRS spectra, extracted from S15.3 pipeline products using the IRS team's SMART program (Higdon et al. 2004), overlaid on Kurucz stellar photosphere models that are minimum $\chi^2$ fit to TD1, Johnson UBVRI, and 2MASS stellar photometry (where available) in Figure 1. Our new reduction improves the signal:noise ratio of the spectra modestly and extends reliable continuum measurements to 38 $\mu$m, beyond which second-order light decreases the SNR. In Table 1, we list the fluxes of our objects in two photometric bands that have been used to search for excess from silicates (8.5-13 $\mu$m) and cold grains (30-34 $\mu$m). The calibration uncertainty in the fluxes is $\sim$5\% and the measured statistical uncertainties are listed in Table 1. We fit the combined IRS, MIPS, and IRAS SEDs using two temperature black body models with the parameters listed in Table 2. We estimate the temperature of the warm component by minimum $\chi^{2}$ fitting the IRS excess at wavelengths shorter than 30 $\mu$m and the temperature of the cool component by estimating the color temperature from the remaining 30-35 $\mu$m and IRAS-60 or MIPS-70 $\mu$m excesses. For HR 8799 with published submillimeter photometry (Williams \& Andrews 2006), cold dust with an emissivity, $\kappa_{\nu}$ $\propto$ $\nu^{\beta}$ where $\beta$ = 1, is needed to fit the far-infrared and submillimeter SED.

Direct imaging of HD 139664 and HR 8799 has provided additional information about their circumstellar environments. High contrast Keck and Gemini AO imaging has detected three planetary mass objects orbiting HR 8799 with projected separations of 24, 38, and 68 AU and inferred masses 7, 10, and 10 $M_{Jup}$, inferred from the measured K-band companion luminosities and the 160 Myr central star isochronal age (Marois et al. 2008). High contrast \emph{HST} ACS scattered light imaging has resolved dust in an edge-on ring around HD 139664 with a possible inner edge at 60 AU, a dust peak at 83 AU, and a sharp outer boundary at 109 AU away from the central star (Kalas et al. 2006). Kalas et al. speculate on the presence of (1) planetary embryos forming at 83 AU that dynamically stir smaller parent bodies, producing a collisional cascade at this radius, (2) a planetary body either inside or outside of the disk that traps migrating dust grains into mean motion resonances, also producing a dusty ring, or (3) planetary bodies inside 60 AU and/or beyond 109 AU that confine dust into the narrow ring.

\section{HR 8799}
HR 8799 was known to possess a debris disk well before the three companions orbiting the central star were discovered. We examine the grain properties to elucidate the planetary system architecture.

We assume that the dust detected in this system via mid- to far-infared thermal emission is gravitationally bound to the system. In this case, we can estimate the minimum grain size by balancing the force due to radiation pressure with the force due to gravity. For small grains with radius, $a$, the force due to radiation pressure overcomes gravity for solid particles smaller than 
\begin{equation}
a_{min} = \frac{3 L_{*}}{8 \pi G M_{*} c \rho}
\end{equation}
where $L_{*}$ is the stellar luminosity. For HR 8799 with $L_{*}$ = 6.1 $L_{\sun}$ and $M_{*}$ = 1.6 $M_{\sun}$ (Chen et al. 2006), we estimate $a_{min}$ = 1.7 and 4.7 $\mu$m, assuming that the grains are either an amorphous carbon/silicate mixture or water ice (densities $\rho$ = 2.5 and 0.91 g cm$^{-3}$), respectively. The 150 K temperature of the warm dust indicates that this population is unlikely to be composed of water ice. The sublimation lifetime of 4.7 $\mu$m water ice grains is expected to be
\begin{equation}
t_{subl} = \frac{a \rho T_{gr}^{1/2} e^{T_{subl}/T_{gr}}}
{\dot{\sigma_{o}}}
\end{equation}
(Jura et al. 1998) where $\dot{\sigma_{o}}$ is the mass rate per surface area (= 3.8 $\times$ 10$^{8}$ g cm$^{-2}$ s$^{-1}$ K$^{1/2}$, $T_{subl}$ = 5530 K; Ford \& Neufeld 2001). For HR 8799, we estimate that 4.7 $\mu$m solid-ice grains sublimate in 40 minutes. Analysis of the 10 $\mu$m and 20 $\mu$m silicate features emitted by T-Tauri disks indicates that the warm dust in these systems is predominately composed of silicates (Sargent et al. 2006). However, silicate grains with $T_{gr}$ = 150 K, are expected to produce a strong 20 $\mu$m silicate emission feature that should be detected using IRS because the dust is optically thin. The lack of a 20 $\mu$m silicate emission feature may indicate that the warm grains may be significantly larger than the minimum grain size and can be modeled assuming that they are large, with emissivities that are constant as a function of wavelength. Another possible reason for the lack of silicate emission features is that the grains are composed of amorphous carbon. Amorphous carbon has a flat emissivity and therefore lacks spectral features. Disentangling the contribution of carbon to this grain composition is challenging. However, since the warm dust component of T-Tauri disks is principally composed of silicates and icy bodies such as comets possess silicate emission features (Min et al. 2005), we believe that the composition in the HR 8799 disk is most likely silicate-rich rather than carbon-rich. The IRAS 60 $\mu$m and SCUBA 850 $\mu$m fluxes can only be modeled using grains with an emissivity that is inversely proportional to wavelength, suggesting that the cold grains are small ($a$ $<<$ $\lambda$/$2 \pi$ = 9.5 $\mu$m), possibly below the sub-blow out size of 1 $\mu$m. 

Black bodies (with 2$\pi a$ $>>$ $\lambda$) in radiative equilibrium around a star with a luminosity, $L_{*}$, and  grain temperature, $T_{gr}$, are expected to possess a distance, $D$, from the central star
\begin{equation}
T_{gr}  =  \left( \frac{L_{*}}{16 \pi \sigma D^2} \right)^{1/4}
\end{equation}
For HR 8799 with $T_{gr}$(warm) = 150 K, we estimate minimum grain distances $D_{in}$ = 8.7 AU, placing the warm dust well within the orbits of the three companions, analogous to asteroidal dust in our solar system.  Small grains (with 2$\pi a$ $<<$ $\lambda$) in radiative equilibrium around a star with a luminosity, $L_{*}$, and  grain temperature, $T_{gr}$, are expected to possess a distance, $D$, from the central star
\begin{equation}
T_{gr}  =  \left( \frac{L_{*} T_{*}}{16 \pi \sigma D^2} \right)^{1/5}
\end{equation}
For HR 8799 with $T_{gr}$(cold) = 40 K, we estimate minimum grain distances $D_{out}$ = 2000 AU, placing the cold dust well outside the orbits of the three companions, analogous to Kuiper belt dust in our solar system. The warm dust temperature is higher than that of water ice sublimation, suggesting that the inner belt may possess rocky grains, while the cold dust temperature is not, suggesting that the outer belt may be icy. 

At the current time, there are few detailed constraints on the orbits of each of the planetary-mass objects. To first order, the system appears face-on with circular orbits because the stellar $v \sin i$ (= 40 km s$^{-1}$) is statistically low for a late A-/early F-type star, the observed orbital motions for the $b$ and $c$ components are orthogonal to their position vectors (relative to the star) and have magnitudes that are commensurate with that expected for face-on circular orbits (Marois et al. 2008). Using the formalism of Chiang et al. (2009), we estimate that the chaotic zone associated with component $b$ extends from 68 AU outward to $\sim$100 AU and that of component $d$ extends inward from 24 AU to 12 AU. The lack of detected dust at distances closer to the orbit of the innermost putative planet may suggest that either there are additional planets in this system that have not yet been detected or that the orbits of the detected planets are slightly eccentric, producing large regions in the disk where circumstellar grains might experience secular perturbations (Moro-Martin et al. 2007). 

Reidemeister et al. (2009) have recently studied comprehensively the properties of the dust and planets around HR 8799, reaching many of the same conclusions; however, they assert that the HR 8799 IRS spectrum may show weak silicate emission and model the cold component using large grains. Their interpretation including a 10 $\mu$m silicate emission feature may arise from the presence of apparent absorption features in the low resolution spectra at 8.5 and 15 $\mu$m that are probably artifacts from the edges of the SL2 and LL2 spectra.

\section{HD 139664}
We can infer the dust architecture in the HD 139664 system assuming that the grains are large and well-described by black bodies. For HD 139664 with $L_{*}$ = 3.6 $L_{\sun}$ and $M_{*}$ = 2.7 $M_{\sun}$ (Chen et al. 2006), we estimate $a_{min}$ = 1.2 $\mu$m (assuming a grain density $\rho_{s}$ = 2.5 cm$^{-3}$). From our SED fit, we measure grain temperatures, $T_{gr}$(warm) = 80 K and $T_{gr}$(cold) = 50 K, corresponding to minimum distances, $D_{in}$ = 21 AU and $D_{out}$ = 61 AU. The similarity between the black body distance for the cold dust grains and the observed scattered light distance suggests that the thermal emission and scattered light are co-spatial, originating from the same grain population. 

We can infer additional grain properties by comparing the scattered light and thermal emission from the cold dust population. We estimate a Bond albedo, $\omega$ = 0.3, assuming $\omega \tau$ = 10$^{-5}$ (Kalas et al. 2006) and $\tau$ = 3.6$\times$10$^{-5}$ from our fit, higher than observed toward the zodiacal dust and the dust in the Fomalhaut disk ($\omega$=0.02, Chiang et al. 2009) but similar to that of $\beta$ Pic moving group member HD 181327 (Chen et al. 2008). We further estimate the ratio of the 0.6 $\mu$m scattering and 70 $\mu$m absorption coefficients. In our simple model, a single population of $N$ grains with radius, $a$, emits a thermal flux,
\begin{equation}
F_{\nu} = \frac{\pi a^2}{d^2} N Q_{abs} B_{\nu}(T_{gr)}
\end{equation}
where $d$ (=17.5 pc) is the distance from the observer to the central star and $Q_{abs}$ is the absorption coefficient. This same grain population is also expected to produce a scattered light surface brightness,
\begin{equation}
SB = \frac{L_{\nu,*}}{4 \pi D^2} f(\theta) Q_{sca} \pi a^2 n
\end{equation}
where $n$ is the grain column density, $f(\theta)$ is the scattering phase function for $\theta$ degrees of light deflection from forward scattering, and $L_{\nu,*}$ and $Q_{sca}$ are the specific stellar luminosity and scattering coefficient at the scattered light wavelength observed (0.5 $\mu$m). We estimate the number of dust grains in the two ansa from the dust column density along our line-of-sight, $n$, assuming a projected area on the sky,  $2 \Delta r \Delta h$, where $\Delta r$ and $\Delta h$ are the measured radial width and the height of the ring, respectively. We derive the following constraint on the scattering and absorption coefficients,
\begin{equation}
\frac{Q_{sca}}{Q_{abs}} = \frac{D^2}{d^4} \frac{B_{\nu}(T_{gr}) SB(gr)}{F_{\nu,*} f(\pi/2)} \frac{2 \Delta r \Delta h}{F_{\nu}(gr)}
\end{equation}
For HD 139664 with $\Delta r$ $\sim$ $\Delta h$ $\sim$ 83 AU, $SB(gr)$ = 11.0 $\mu$Jy arcsec$^{-2}$ (Kalas et al. 2006) and $F_{\nu,*}$ = 52.6 Jy at V-band, $Q_{sca}(0.5 \mu m)/Q_{abs}(70 \mu m)$ = 7.3$\times$10$^{-4}$/f($\pi$/2). Typically, $Q_{sca}/Q_{abs}$ is of order unity for large grains, suggesting that the scattering phase function at 90$\arcdeg$ scattering, $f(\pi/2)$, must be very small. Assuming the Henyey-Greenstein phase function, we can rewrite the constraint on the scattering and absorption coefficients as a function of $g$, the asymmetric scattering parameter, $f(\pi/2) = (1-g)^2/4 \pi (1+g)^{3/2}$ (Henyey \& Greenstein 1941) and search for materials that satisfy this constraint. We calculate scattering and absorption coefficients and asymmetric scattering parameters using laboratory measurements of the optical constants for amorphous olivine (MgFeSiO$_{4}$; Dorschner et al. 1995) solid and fluffy grains (assuming a vacuum volume fraction of 0.9) using Bruggeman effective medium and Mie theory (Bohren \& Huffman 1983). We find that large ($a$ $>$ 40 $\mu$m), porous olivine grains can satisfy our grain properties constraint while solid grains can not. Large, porous grains have been used to explain the strong forward scattering and high polarization fraction observed in the AU Mic disk using \emph{HST} ACS (Graham et al. 2007).

\section{$\lambda$ Boo}
The star $\lambda$ Boo has become the archetype of a class of Population I late B- to early F-type stars with moderate to extreme (up to a factor of 100) surface underabundances of Fe-peak elements and solar abundances of lighter elements (C, N, O, and S) (Paunzen 2004). One possible hypothesis for the abundance anomalies seen toward $\lambda$ Boo stars is that they selectively accrete material from their circumstellar environments. In this scenario, grains with icy mantles and refractory cores slowly spiral in under Poynting-Robertson drag, sublimating or photo-desorbing water ice as they migrate inward. Once their volatile mantles have been depleted, their refractory cores are expelled via radiation pressure and the gas is accreted onto the star. This hypothesis predicts that $\lambda$ Boo stars should possess debris disks with constant surface densities, governed by Poynting-Robertson Drag, and SED's with $F_{\nu}$ $\propto$ $\nu^{-1}$ (Jura et al. 2998). Preliminary SED analysis of 6 debris disks around A-type stars with IRS excess indicated the possibility that two $\lambda$ Bootis stars ($\lambda$ Boo and HR 1570) possessed disks with with $F_{\nu}$ $\propto$ $\nu^{-1}$ and therefore uniform surface densities in contrast to debris disks around four normal A-type stars that possessed disks with cleared inner regions, better fit by single temperature black bodies (Jura et al. 2004). At the time, this observational result was consistent with the idea that the abundance anomalies of $\lambda$ Boo stars might be generated by pollution of their stellar atmospheres via selective accretion of volatiles from inward migrating circumstellar dust grains. Since this result was published, more data have been obtained and data reduction of spectra obtained early in the \emph{Spitzer} mission has improved.

The connection between $\lambda$ Boo stars and circumstellar and/or interstellar dust is highly uncertain today. Su et al. (2008) carried out a MIPS 24 and 70 $\mu$m survey searching for infrared excess associated with $\sim$160 A-type stars, including 15 $\lambda$ Boo stars (R. Gray's $\lambda$ Boo website: http://www1.appstate.edu/dept/physics/spectrum/lamboo.txt) with an average age of $\sim$300 Myr. Their observations suggest that the 24 and 70 $\mu$m excess rate for $\lambda$ Boo stars is higher than average, even though the average age for the aggregate sample is similar $\sim$270 Myr. We estimate that 8/15 (53\%) and 7/9 (78\%) of the $\lambda$ Boo stars possess 24 and 70 $\mu$m excesses, respectively, compared with 53/160 (33\%) and 39/69 (57\%) of the total sample, suggesting that the $\lambda$ Boo phenomenon may be associated with circumstellar dust. Low resolution IRS spectra of 6 $\lambda$ Boo stars have been published thusfar: HD 30422, HR 1570, HD 74873, HD 110411, $\lambda$ Boo, and HD 183324. New models for the IRS spectra of HD 30422 and HD 110411 indicate that these objects are better fits using a power law, consistent with a uniform surface density  disk with an inner radius coincident with the stellar radius (Morales et al. 2009); however, revised models for $\lambda$ Boo and HR 1570 (Chen et al. 2006) spectra and newly published models for HD 74873 and HD 183324 (Morales et al. 2009) suggest that these objects are better fit using a black body, suggesting that the disks around these stars possess central clearings. Therefore, the mechanism by which grains material is selectively accreted may be diverse and may need to be revised. In any case, none of the objects presented here can be modeled using the selective accretion hypothesis because their SEDs are not well-fit assuming $F_{\nu}$ $\propto$ $\nu^{-1}$.

\section{Discussion}
At the present time, a dozen systems are known to possess multiple dust belts, suggesting the presence of multiple parent body populations, perhaps analogous to the asteroid and Kuiper belts in our Solar System (See Table 2). These dusty bands apparently occur around stars with a wide range of spectral types and ages. Some are identified using SED modeling while others are discovered via high resolution imaging. High resolution imaging has discovered unresolved excesses (consistent with asteroidal dust) around objects whose SEDs are well fit using a single cold dust population, indicating that multiple dust belts may be common in debris disks. The Spitzer Formation and Evolution of Planetary Systems (FEPS) team surveyed 328 solar-like stars with ages 0.003 - 3 Gyr quantifying the evolution of dust and gas around main sequence stars using IRAC, IRS, and MIPS, finding that $\sim$10\% solar-like stars posses cold debris disks. Of the dusty disks, fully 1/3 are apparently better fit using multiple temperature components than single temperature black bodies (Hillenbrand et al. 2008), also suggesting that extended disks and multiple dust components may be common in debris disks.

The present day asteroid and Kuiper Belts possess an estimated 0.0005 and 0.01 $M_{\earth}$ in large bodies, respectively, significantly smaller than originally inferred to exist in these regions ($\sim$2-3 and $\sim$35 $M_{\earth}$, respectively). The migration of the outer planets and the subsequent crossing of the 2:1 resonance by Jupiter and Saturn is believed to have ejected or scattered the majority of the initial mass in the asteroid and Kuiper Belts (Levison et al. 2007). We estimate the mass for each debris belt observed toward $\lambda$ Boo, HD 139664, and HR 8799 (see Table 2) assuming that the current mass in parent bodies is at least as large as the amount of dust that has been removed from the system thus far, $M_{PB}$ $\geq$ 4 $L_{IR} t_{age}/c^{2}$ (Chen \& Jura 2001), assuming stellar ages, $t_{age}$ = 310, 300 (Kalas et al. 2006), and 160 Myr (Marois et al. 2008), respectively.  These estimates are a lower bounds because radiation pressure (and not Poynting-Robertson Drag) probably dominates the grain removal process. We find that (1) the minimum parent body masses in the warm and cold belts are 0.0002-0.003 $M_{\earth}$ and 0.002-0.004 $M_{\earth}$, respectively, similar to the present day asteroid and Kuiper belts and (2) the cold dust belts are 1-10$\times$ more massive that the warm ones, suggesting that they have more similar masses that the present day asteroid and Kuiper belts.  

\section{Conclusions}
We have rereduced the IRS spectra of three debris disks  and reanalyzed their SEDs. We conclude the following:

1. The infrared SEDs of $\lambda$ Boo, HD 139664, and HR 8799 are well-fit using two single temperature black bodies indicating the presence of two dust populations. For HR 8799, the warmer dust is interior to the newly discovered orbiting planets in an exo-zodiacal dust belt and the cooler dust is exterior to the planets in an exo-Kuiper belt.

2. The cool grains in the HD 139664 must be strongly, forward scattering, consistent with large ($a$ $>$ 40 $\mu$m), porous grains if it produces both the thermal emission detected using \emph{IRAS} and the scattered light disk  resolved using \emph{HST} ACS at visual wavelengths.

3. The debris disks around the majority of observed $\lambda$ Boo stars apparently possess inner clearings similar to those around normal A-type stars, suggesting that simple dusty disks delivering grains from the outer regions of disks onto central stars are not responsible for the peculiar stellar abundances observed.

\acknowledgements
We would like to thank J. Debes, M. Fitzgerald, W. Forrest, J. Graham, M. Jura, B. Macintosh, and J. Patience for their helpful comments and suggestions.

\begin{deluxetable}{lcccc}
\singlespace
\tablecaption{Measured Fluxes}
\tablehead{
    \colhead{Name} &
    \colhead{Spectral} &
    \colhead{Distance} & 
    \colhead{F$_{\nu}$(8.5-13 $\mu$m)} &
    \colhead{F$_{\nu}$(30-34 $\mu$m)} \\
    \omit &
    \colhead{Type} &
    \colhead{(pc)} &
    \colhead{(mJy)} &
    \colhead{(mJy)} \\
}
\tablewidth{0pt}
\tablecolumns{5}
\startdata
$\lambda$ Boo & A0p & 30 &  934$\pm$5 & 258$\pm$7  \\
HD 139664 & F5IV-V & 18 & 1256$\pm$7 & 211$\pm$5 \\
HR 8799 & A5V & 40 &  289$\pm$3 & 65$\pm$6 \\
\enddata
\end{deluxetable}

\begin{deluxetable}{l|ccccc|ccccc|}
\singlespace
\tablecaption{Dust Properties} 
\tablehead{
    \omit &
    \multicolumn{5}{c}{Warm Dust} &
    \multicolumn{5}{c}{Cold Dust} \\
    \colhead{Name} &
    \colhead{$\beta$} &
    \colhead{$T_{gr}$} & 
    \colhead{$D$} & 
    \colhead{$L_{IR}/L_{*}$} &
    \colhead{$M_{PB}$} &
    \colhead{$\beta$} &
    \colhead{$T_{gr}$} &
    \colhead{$D$} & 
    \colhead{$L_{IR}/L_{*}$} &
    \colhead{$M_{PB}$}\\
    \omit &
    \omit &
    \colhead{(K)} &
    \colhead{(AU)} &
    \omit &
    \colhead{$M_{\earth}$} &
    \omit &
    \colhead{(K)} &
    \colhead{(AU)} &
    \omit &
    \colhead{$M_{\earth}$} \\
}
\tablewidth{0pt}
\tablecolumns{11}
\startdata
$\lambda$ Boo & 0 & 110 & 27 & 5.8$\times$10$^{-6}$ & 0.003 & 0 & 60 & 115 & 7.6$\times$10$^{-6}$ & 0.004 \\
HD 139664 & 0 & 80 & 21 & 4.2$\times$10$^{-6}$ & 0.0004 & 0 & 50 & 61 & 2.4$\times$10$^{-5}$ & 0.003 \\
HR 8799 &  0 & 150 & 9 & 1.9$\times$10$^{-6}$ & 0.0002 & 1 & 40 & 2000 & 2.5$\times$10$^{-5}$ & 0.002 \\
\enddata
\end{deluxetable}


\begin{deluxetable}{lccccccc}
\singlespace
\tablecaption{Multiple Dust Belt Systems} 
\tablehead{
    \colhead{HR} &
    \colhead{HD} &
    \colhead{Name} &
    \colhead{Spectral} &
    \colhead{Age} & 
    \colhead{$D_{hot}$} & 
    \colhead{$D_{cold}$} &
    \colhead{References} \\
    \omit &
    \omit &
    \omit &
    \colhead{Type} &
    \colhead{(Myr)} &
    \colhead{(AU)} &
    \colhead{(AU)} &
    \omit \\
}
\tablewidth{0pt}
\tablecolumns{8}
\startdata
1084 &  22049 & $\epsilon$ Eri    & K2V     & 850 & $\sim$3 & 35-90 & 1 \\  
2020 &  39060 & $\beta$ Pic         & A6V     & 12 & $<$5 & 100-1500 & 10, 11 \\
3927 &  86087 &                              & A0V     & 50 & $\sim$7 & $\sim$50 & 2 \\
4775 & 109085 & $\eta$ Crv         & F2V      & 1000 & $\sim$2 & 130-170 & 2, 12 \\
4796 & 109573 &                             & A0V      & 8 & $<$20 & 60-87 & 5, 8 \\
          & 113766 &                              & F3/F5V & 16 & $\sim$4 & $\sim$30-80 & 2 \\
5351 & 125162 & $\lambda$ Boo & A0p      & 310 & $\sim$24 & $\sim$115 & this work \\
          & 139664 &                              & F5IV-V & 200 & $\sim$15 & 60-110 &  4, this work \\
7329 & 181296 & $\eta$ Tel           & A0Vn   & 12 & $\sim$5 & $\sim$30 & 2 \\
8728 & 216956 & Fomalhaut          & A4V     & 200 & $<$10 & 130-160 & 3, 9 \\
8799 & 218396 &                              & A5V     & 160 & $\sim$8    & $\sim$2000 & 7, this work \\
\enddata
\tablerefs{(1) Backman et al. 2009; (2) Chen et al. 2006; (3) Kalas et al. 2005; (4) Kalas et al. 2006; (5) Koerner et al. 1998; (6) Lisse et al. 2008; (7) Marois et al. 2008; (8) Schneider et al. 2009; (9) Stapelfeldt et al. 2004; (10) Telesco et al. 2005; (11) Wahhaj et al. 2003; (12) Wyatt et al. 2005}
\end{deluxetable}


\begin{figure}
\plottwo{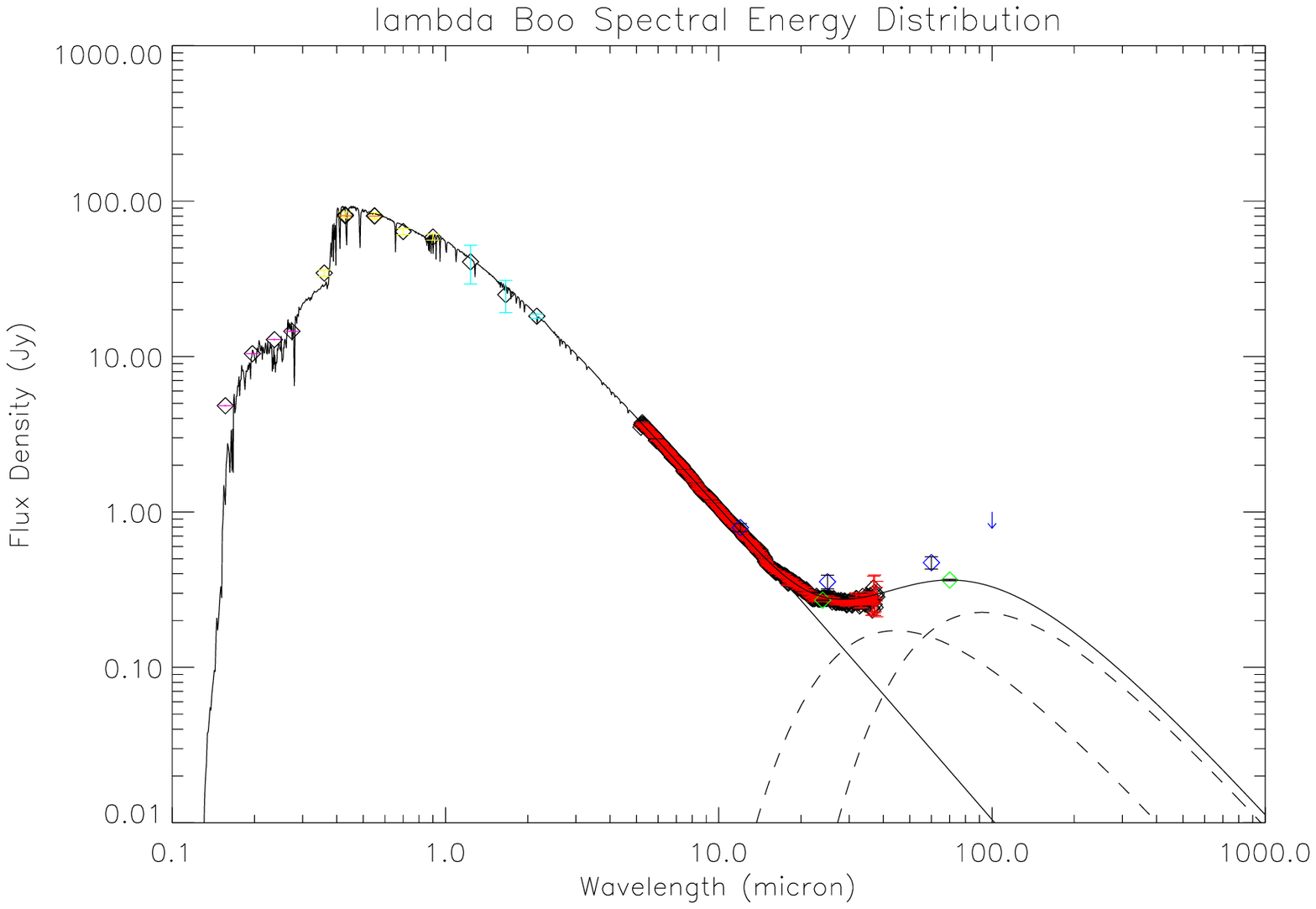}{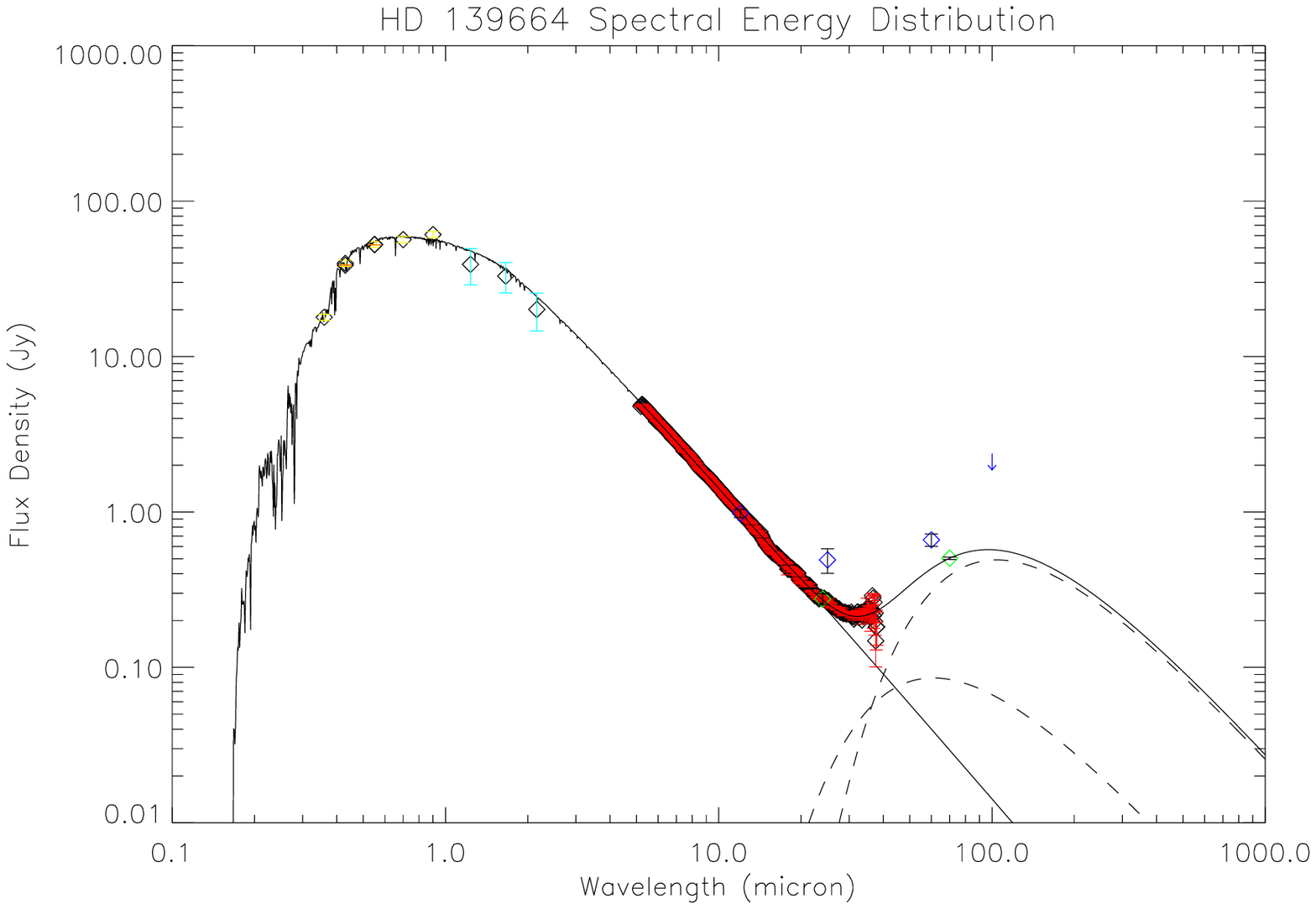}
\plotone{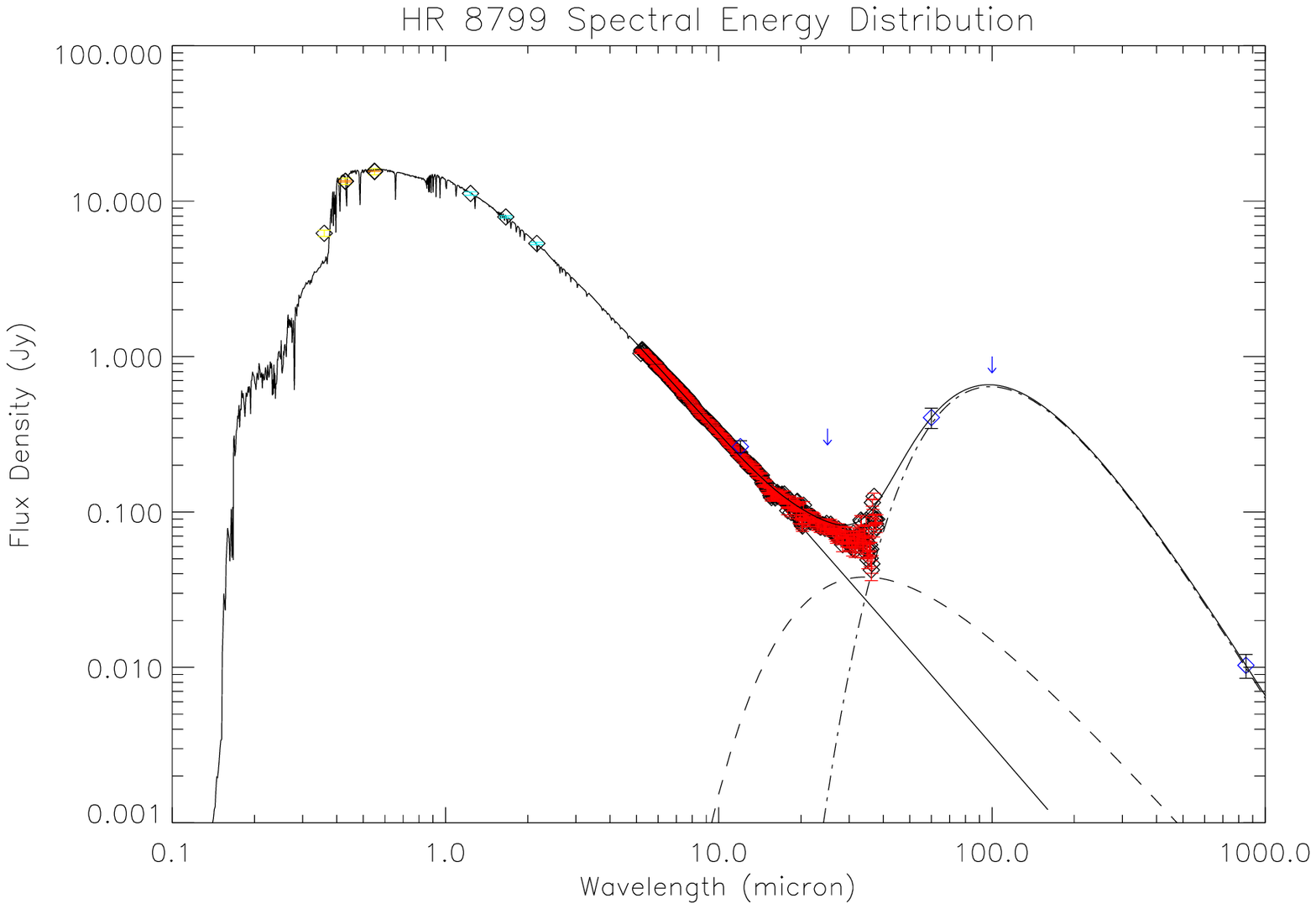}
\caption{SEDs for $\lambda$ Boo, HD 139664, and HR 8799: TD1 fluxes are plotted with magenta error bars; General Catalogue of Photometric Data mean UBV or Johnson et al. fluxes are plotted with yellow error bars; and 2MASS JHK fluxes are plotted with cyan error bars. Color-corrected IRS, MIPS, and IRAS data, where available, are shown with red, green, and blue error bars.}
\end{figure}

\begin{figure}
\plottwo{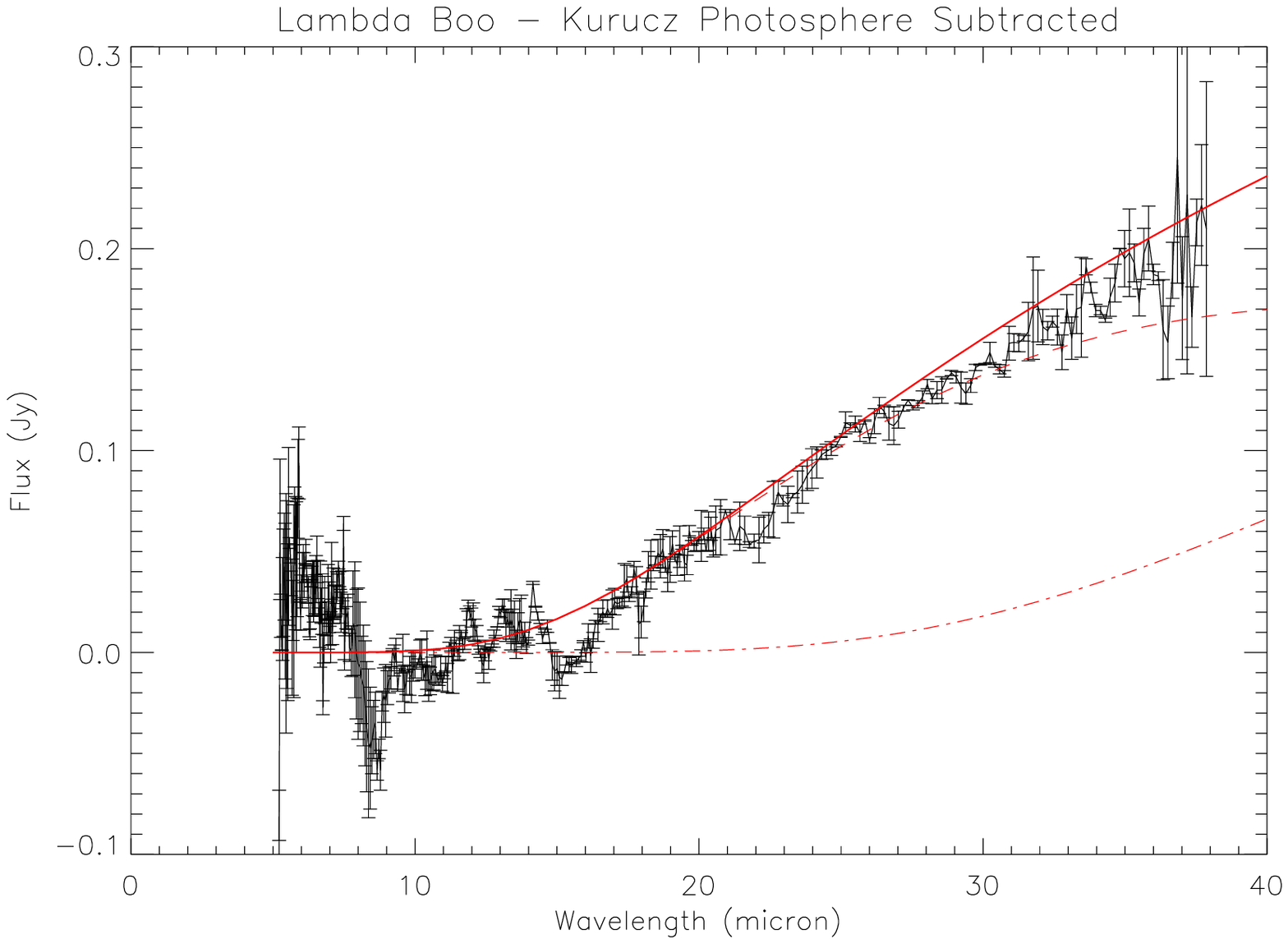}{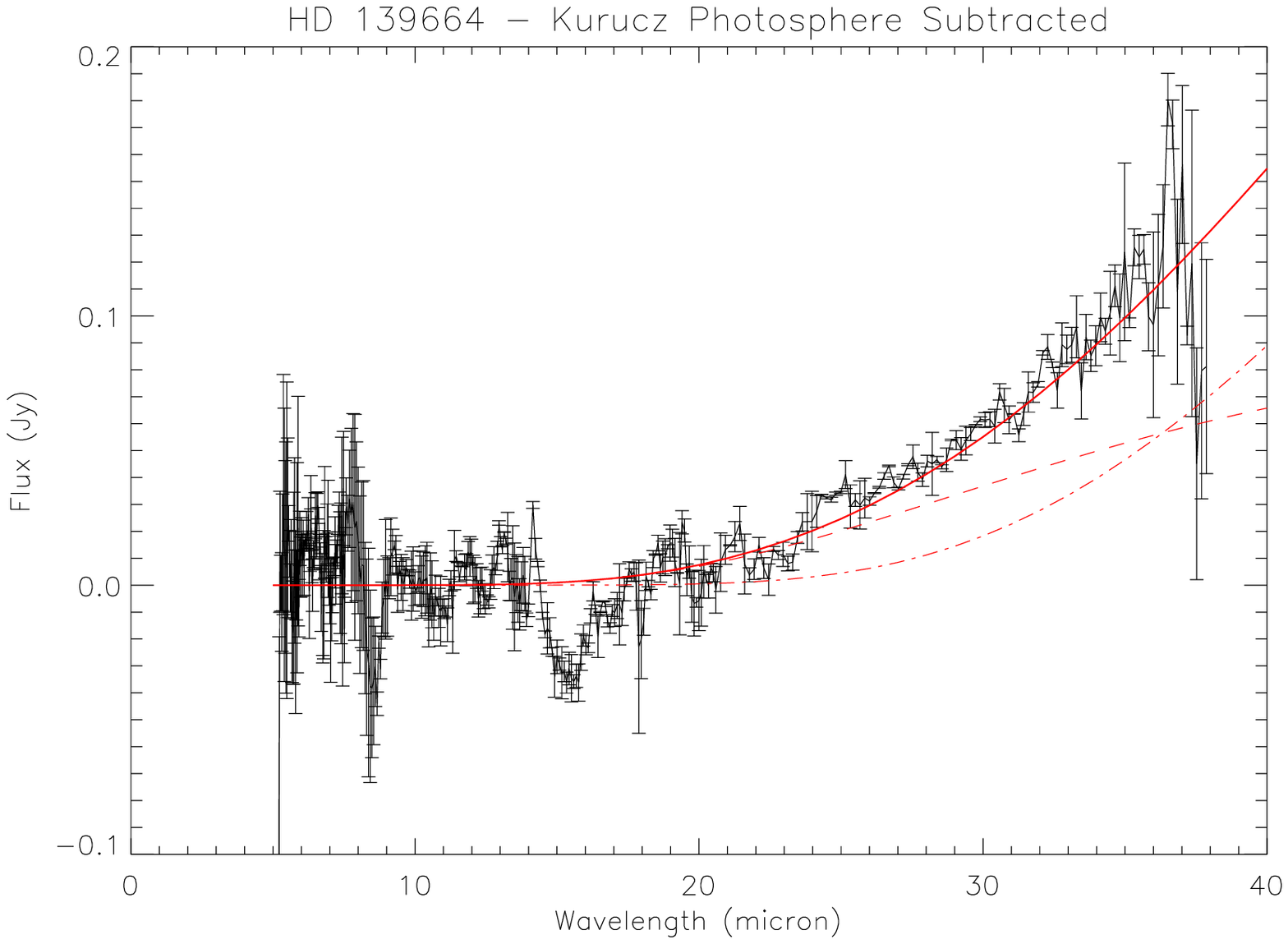}
\plotone{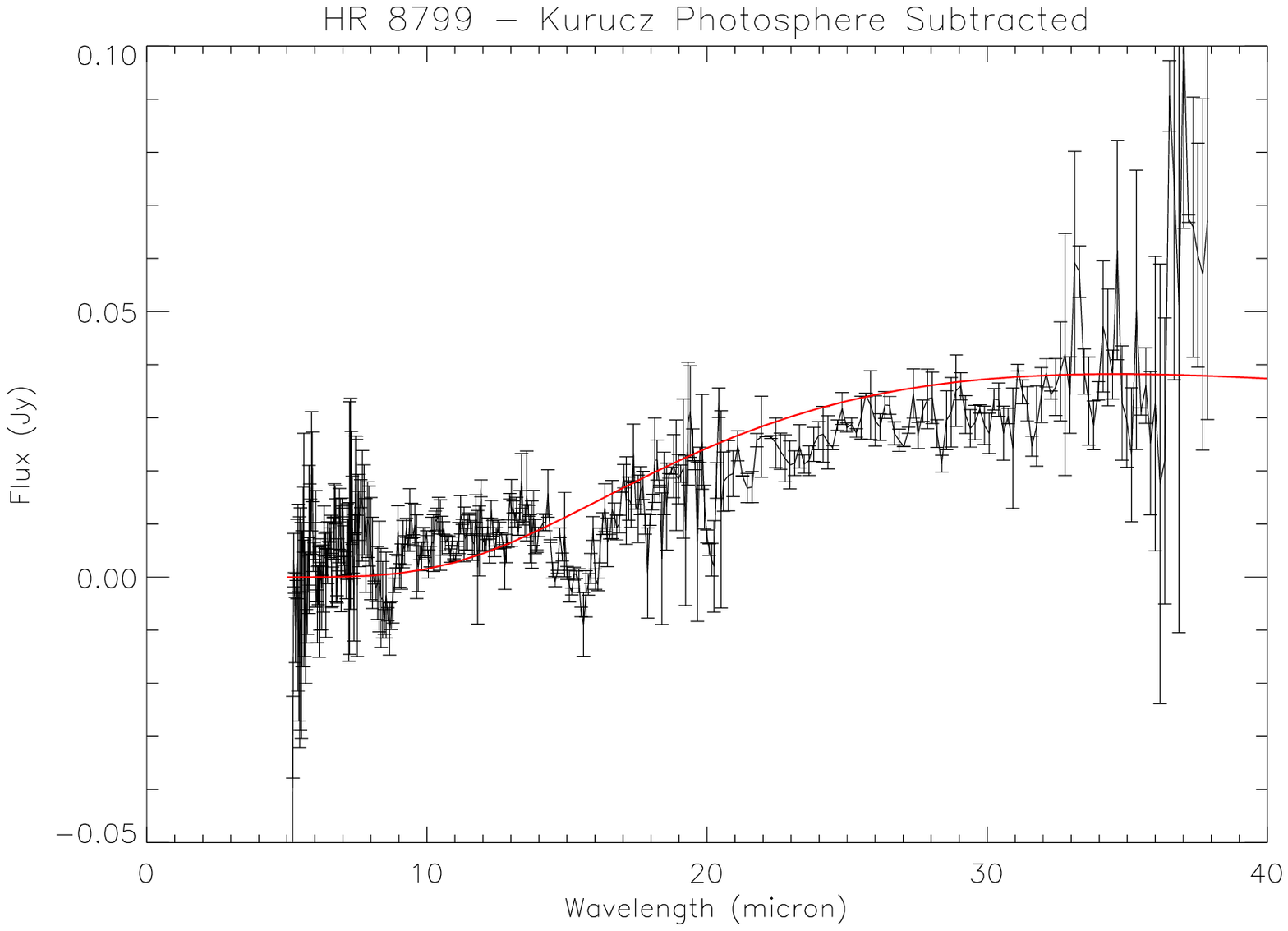}
\caption{Photosphere-subtracted IRS spectra with $F_{\nu}$ plotted as a function of wavelength. The minimum $\chi^2$ fits for the two single temperature black body component are overlaid with the warm dust component shown as a dashed line, the cold dust component shown as a dashed-dotted line, and the sum of the two component as a solid line.}
\end{figure}

\end{document}